\documentstyle[twoside,ijmpc1]{article}

\catcode`\@=11
\long\def\@makefntext#1{
\protect\noindent \hbox to 3.2pt {\hskip-.9pt  
$^{{\eightrm\@thefnmark}}$\hfil}#1\hfill}               

\def\@makefnmark{\hbox to 0pt{$^{\@thefnmark}$\hss}}    

\def\ps@myheadings{\let\@mkboth\@gobbletwo
\def\@oddhead{\hbox{}
\rightmark\hfil\eightrm\thepage}   
\def\@oddfoot{}\def\@evenhead{\eightrm\thepage\hfil
\leftmark\hbox{}}\def\@evenfoot{}
\def\sectionmark##1{}\def\subsectionmark##1{}}

\def\qed{\hbox{${\vcenter{\vbox{                        

   \hrule height 0.4pt\hbox{\vrule width 0.4pt height 6pt
   \kern5pt\vrule width 0.4pt}\hrule height 0.4pt}}}$}}

\def\bsc{{\sc a\kern-6.4pt\sc a\kern-6.4pt\sc a}}       

\def\bflatex{\bf L\kern-.30em\raise.3ex\hbox{\bsc}\kern-.14em 
T\kern-.1667em\lower.7ex\hbox{E}\kern-.125em X}

\newcommand{\mboxf}[1]{\framebox(16,16){\large #1}}

\begin{document}
\runninghead{F. Babalievski}{Hoshen-Kopelman algorithm vs. spanning
  tree ...}
\normalsize\textlineskip
\thispagestyle{empty}
\setcounter{page}{1}
{\footnotesize Submitted to Intern. J. of Modern Physics C}

\vspace*{0.88truein}
\fpage{1}
\centerline{\bf CLUSTER COUNTING: THE HOSHEN-KOPELMAN}
\vspace*{0.035truein}
\centerline{\bf ALGORITHM vs. SPANNING TREE APPROACHES}
\vspace*{0.037truein}
\centerline{\footnotesize F. BABALIEVSKI\footnote{Permanent address:
    Institute of General and Inorganic Chemistry, 
     Bulgarian Academy of Sciences, 1113 Sofia, Bulgaria}}
\vspace*{0.015truein}
\centerline{\footnotesize\it Institute for Computer Applications 1,
  University of  Stuttgart}
\baselineskip=10pt
\centerline{\footnotesize\it  Stuttgart, 70569, Germany}

\vspace*{0.25truein}

\abstracts{Two basic approaches to the cluster counting task
in the percolation and related models are discussed. The
Hoshen-Kopelman
multiple labeling technique for cluster statistics is redescribed.
Modifications for random and aperiodic lattices are sketched as well
as some parallelised versions of the algorithm are mentioned.
The graph-theoretical basis for the spanning tree approaches
is given by describing the {\em breadth-first search} and {\em depth-first
search} procedures. Examples are given for extracting the elastic
and geometric ``backbone'' of a percolation cluster. 
 An implementation of the ``pebble game'' algorithm using
 a depth-first search method is also described.}{}{}

\textlineskip
\vspace*{12pt}
\noindent 
The algorithmic task of performing cluster statistics (i.e. cluster  
counting) is almost as old as the computer calculations: one of
the pioneers of  percolation theory, S.~Broadbent, presented 
a percolation problem$^1$
at one of the first conferences on 
Monte Carlo 
simulations back in 1954. Since then  for the physics community
 the cluster statistics problem has been tightly connected
with the percolation theory.$^{2,3}$ 
Another co-founder of
this theory, J.~M.~Hammersley,  argued in the 60-s that the 
numerical study of percolation models  is strongly 
hindered by the lack of fast enough computers. He predicted that this 
problem will remain even for the computers of the next century. 
Only a few years later
Hoshen and Kopelman proposed$^4$
their cluster-labeling algorithm 
(further quoted as HK76),
which made the pessimistic view of J. Hammersly completely irrelevant. 
Before HK76 it was
believed that the computational efforts grow faster than the squared number of
``particles'' subject to clustering. (Particle here means any geometrical 
object which could be connected under certain rules with some of its 
neighbors.)
The HK76 algorithm proved that this relationship can be linear.  That was the 
real breakthrough, because very often, especially for percolation models,
the number of particles might be well beyond $10^6$: the world largest model
percolation system being simulated, has (more than)  $10^{14}$
particles. 
Moreover, the HK76 algorithm solved the serious problem of lack of
computer memory for very large percolation systems.

Next we will describe the cluster counting problem itself, then the HK76
algorithm and some modifications will be presented. Afterward the 
graph-theoretical
framework for cluster descriptions will be sketched. The discussion section 
will be on some comparisons and combined approaches.


\section{The cluster counting}

Remaining 
on the ground of  physics intuition, let us imagine that
we have a large number of objects spread somehow in space. Such objects could
be atoms, monomers, polymers, sand grains, ... telephones, computers, 
computer networks, .. stars, galaxies. For each pair
of objects we may determine (following certain rules -- often probabilistic) 
whether they are directly
connected (i.e. adjacent, bonded) or not. Indeed some objects might be indirectly connected via chain(s) 
of direct connections. 

 Each set of interconnected particles forms a (connectivity) cluster.
For each (random) configuration, we will need to extract some statistics for 
the distribution of the clusters -- their size, form, fractal dimension.
Then a statistics for an ensemble of configuration could be made.

So, the {\em cluster counting} could be thought as  the 
following task:

\begin{enumerate}
\item Create a (computer) model structure of the objects for which the
  spread of
connectivity will be studied. In some cases that could be a digitized image
of natural objects, e.g. a colony of bacteria.

\item Decide, for each pair of objects, if they are adjacent (bonded) or not.

\item Identify the clusters of connected objects.

\item Make statistics of certain properties of the clusters.
Store the statistics for further use.

\item Repeat  1 -- 4 (or 2 -- 4)  enough times in order to have reliable 
counts for the statistics made within each single realizations;  or,
follow the evolution of the cluster statistics with time, if the
realizations are not independent.
\end{enumerate}

The difficulties are concentrated mainly in item 3 of the list above
and the present paper
focuses on it. The other items could be also nontrivial as well, and what 
follows is a short discussion about them.

Item 2 might be a problem (mostly in non-physics applications)
if there is no {\em a priori}\/ information about which pairs of
objects are bonded.
Instead  there could be some "plausible" assumptions. For example, 
when the objects are 
situated in two dimensions one could presume that the bonds
should form a mosaic (would not cross each other) and 
among all possible mosaics  one should chose the ``less heavy'' mosaic. One 
attaches 
to each bond (edge of the mosaic) a certain weight (e.g. its length, or the
square of the length), and then searches for the mosaic with a minimal weight. 
It turns out that this search is analogous to the problem of
constructing a minimum spanning tree in the framework of graph theory: 
a further discussion can be found in $^5$
and references therein.

If the number of objects is very
large, the algorithmic bottleneck might be item 1. This problem
could be alleviated using techniques similar to these
applied in calculation with very large(``monster'') 
matrices,$^{6,7}$ 
but this fact does not decrease the virtue of HK76 method to use 
$d-1$ dimensional cuts of the ``sample'' -- see the next section....

\section{The Hoshen-Kopelman Method and its modifications}

As many seminal methods, the HK76 algorithm works best for the example,
on which it was demonstrated first -- the ordinary percolation model on
square lattice. Let us describe first the  bond percolation model itself.

The ``objects'', we have discussed up to now about, are, in the 
present example, the sites of a square
cut (with size $L$) from the infinite square lattice. 
Each lattice bond might be ``open'' with probability $p$ ($0 \le p \le 1$).
As it is known, there is a critical probability $p_c$ such that 
 for  $p \ge p_c$ a spanning cluster of open bonds (a cluster which
connects the edges of the lattice) appears with probability $S(L,p)$ and
$\lim_{L \rightarrow \infty} S(L,p) = 1$.

We will describe the HK76 algorithm using an analogy with a 
``brick layering'' in which each lattice site is a brick in the wall
(the lattice).
The ``brick layering'' (for given $p$ - in this analogy this is the
probability
to put mortar between two bricks)
starts,  say, from the lower left brick (vortex) of the wall (the lattice). One
adds from left to right vortex after vortex, each time checking if the new 
vortex is connected with the last  one. This ``check'' consists simply in 
drawing a (pseudo) random number distributed equally between $0$ and $1$: if 
the number is less than $p$ the vortices become bonded. 

 We start describing  the construction of the first row:
We assign to the   left most
vortex a label:\/   "1". If the next vortex added in the row is
connected with the first it accepts the same label ("1").
The label "2" is assigned to the first vortex inside the 
row for which the connectivity check fail. In general; if the next 
vortex appears connected to its previous neighbor it accepts 
the same label --- if not, it accepts the next number as a label. 
So the vortices with the same labels form (one-dimensional) clusters.

Contrary to the masons' practice, {\em the second row} is created by adding
bricks in a way so that they contact only  one brick from the lower
layer. This time the connectivity check should be made with two neighbors:
the previous vortex in the same row and the lower vortex.

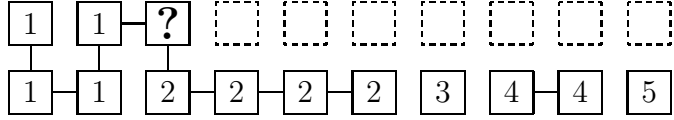
\begin{figure}
\label{ClConflict}
\caption{Cluster identification by labeling: a labels conflict}
\begin{picture}(300,75)(-40,1)
\put(0,10){\mboxf{1}}      \put(16,18){\line(1,0){10}}
\put(26,10){\mboxf{1}}
\put(52,10){\mboxf{2}}     \put(68,18){\line(1,0){10}}
\put(78,10){\mboxf{2}}     \put(94,18){\line(1,0){10}}
\put(104,10){\mboxf{2}}    \put(120,18){\line(1,0){10}}
\put(130,10){\mboxf{2}}
\put(156,10){\mboxf{3}}
\put(182,10){\mboxf{4}}    \put(198,18){\line(1,0){10}}
\put(208,10){\mboxf{4}}
\put(234,10){\mboxf{5}}

\put(0,36){\mboxf{1}} 
     \put(8,26){\line(0,1){10}}
\put(26,36){\mboxf{1}}    \put(42,44){\line(1,0){10}}
      \put(34,26){\line(0,1){10}}
\put(52,36){\framebox(16,16){\LARGE\bf ?}}
      \put(60,26){\line(0,1){10}}
\multiput(78,36)(26,0){7}{\dashbox{2}(16,16){ }}
\end{picture}
\end{figure}

Here  a cluster label conflict may appear:  the new vortex bonds
 with both neighbors, and  both may have different labels
 attached  (See Fig.\ref{ClConflict} and the similar figure in ...).
Indeed, there is
a  ``brute force'' method for solving it: attach to the new vortex the smaller 
of  the two numbers (e.g. $n_1 < n_2$); then change all labels "$n_2$" to
labels "$n_1$" -- this  is the method that made Hammmersley so pessimistic.

The  HK76 approach avoids this need for backward relabeling. 
Instead of changing the vortex's labels, a one-dimensional array of 
``labels-of-labels'' ($LL[..]$) is created. The size
of this array should be greater than the possible number of different labels
which could be needed. Then the i-th element of this array is ``the label'' 
of all labels "i" put on vortices during the ``brick layering procedure''. 
This time a cluster may have vortices with different labels.

The HK76 method  needs one more variable:
it stores the number of the last used label. When a new vortex ("V") is added 
and it is not bonded to its two neighbors this variable increases by one
and the vortex label obtains its value (say $j$). 
 The respective element of the 
``labels-of-labels'' is set to unity ($LL[j]:=1$) -- the ``current size'' of 
the cluster
to which the new vortex belongs. If there is no label conflicts,
every new vortex, which is a direct or indirect neighbor to the vortex "V",
accepts the same label and the  content of
$LL[j]$ is increased: $LL[j]:=LL[j] +1$. 

 If a label conflict occurs then the label of the new vortex 
 accepts the smaller
 number ($n_1$ as above), but this time there is no other change of labels.
 The values of $LL[n_2]$ and $LL[n_1]$ change instead. If the label conflict 
 occurs for the first time, this change is simply:
 
 \begin{displaymath}
 LL[n_1]:=LL[n_1]+LL[n_2]; \;\;\; LL[n_2]:= - n_1
 \end{displaymath}

In that way, the size of the cluster is stored in the array element $LL[n_1]$ 
and
$LL[n_2]$ points to the ``root'' of the cluster, or, the cluster ``owner''. 
When one examines the third or higher row, the summation above could be
done only after checking if the values of $LL[n_1]$ and $LL[n_2]$ are positive.
If not, the chain of pointers should be followed:
$$
{\bf while}\;\; LL[n_i] < 0\;\; {\bf do} \;\; n_i:=-LL[n_i]; $$

in order to find the current ``root'' for that cluster.
The details of this procedure could be found elsewhere ---
e.g.$^2$  
 the original paper$^4$ 
or the very recent paper of Hoshen et 
al.$^8$

If we return to our brick layering, we can see that the considerations for the 
first two rows could be repeated for the $m$-th and $(m-1)$-th layers.
Without entering into details, the algorithm  sketched above allows one to use
only two layers during the simulation. After completing
the lattice sweep (i.e. "the brick layering"), the information needed
for cluster size statistics is stored
in the array of ``labels-of-labels'' -- in its positive elements.  

The  HK76 algorithm has to be modified in order to allow the identification
of a spanning cluster 
--- a cluster which connects two opposite edges of the 
lattice. This could be done if label "1" is ascribed to the vortices
in the first column and label "2" to the right-most column.

Indeed, properties of clusters other than their size and their spanning
could be of interest: e.g.  the radius of gyration, the shape anisotropy,
certain fractal dimensions and so on; all that for the cluster itself or for 
subsets of its vortices and/or bonds.

Usually the first thing to be done is to ``extract`` a certain cluster -- 
that is to find the
coordinates of all vortices which belong to it. If the whole structure is 
kept in memory, this task can just be accomplished  by inspection 
of the $LL[..]$
array. The inspection shows which labels (attached to certain
vortices) corresponds to  
the same cluster, even if there could be multiple labels for one cluster.

In the original HK76 method, the random percolation structure is created
``on the fly'' -- during the ``brick layering'', as mentioned above. 
This portion of the percolation structure which remains behind the
``growth zone'' is destroyed, so one keeps in the memory only a small
fraction ( $\propto (d-1) L$, where $d$ is the spatial dimension) of the
percolation configuration. After completing the ``layering''
one gets the whole set of labels which form
separate clusters. But the coordinates of the respective vortices are
lost at that moment!

A way out is to keep the whole percolation structure in memory, and to lose
the memory saving advantage of HK76. Instead, the $(d-1) L$ memory cost might 
be preserved as follows:$^9$

 After the first ``brick layering'' for a certain configuration, one memorizes
 the array $LL[..]$ (say, as $LLold[..]$). Then the procedure is repeated
 with {\em the same} pseudo-random number sequence. During the second
 sweep, after attaching a label to a new vortex one can determine   to 
 which cluster belongs this vortex. This can be done by 
 examination of the $LLold[..]$ array.
 
Indeed, modifications of HK76 algorithm for other periodic lattices are 
more or less straightforward. The case of non periodic lattices needs 
some discussion and will serve as a bridge to the Graph-Theoretical
approaches described in Sec.\ref{G-TA}.

\subsection{A modification for non periodic lattices}

We may return to the brick layering analogy and two dimensional structures. 
The following consideration 
could be easily  applied to dimensions higher than two.

The space region, where the connecting objects are situated, is divided into
sub-cells -- so called ``imaginary covering mesh''.$^{10,11}$
Then 
the brick layering consist in adding ``bricks'' with the
size of a sub-cell. Each brick may ``contain'' one or more objects. 
The number of objects depends on the specific problem: the distribution of 
the objects' shape and size, the connectivity rules and so on.

We will refer here to a  soft-shell-hard-core disks continuum
percolation model (named also {\em penetrable concentric shell}\/ 
model$^{12}$).

One deals with randomly spread disks with equal radii (say, $r=1$). 
The disks are  ``penetrable''
up to a distance $\lambda r$ from their center ($\lambda \in [0,1]$).
Two disks ``are connected'' if the distance between their centers is
less than $r$ (and greater than $\lambda r$). Indeed the hard core restrictions
lead to correlations which can differ depending on the way the random structure
was created.

The limit $\lambda = 0$ leads to the freely overlapping disks model,
and the opposite limit ($\lambda =1$) gives the hard disks model. 

The use of space sub-cells and the  ``bricks'' depends on the value 
of $\lambda$.
For $\lambda$ not very small (e.g. $\lambda > 0.1$), one can prescribe that no
more than one disk center can lie in a sub-cell. Then the sub-cells should be
squares with edge length (slightly) less than $\sqrt{ \frac{\lambda r}{2}}$ 
  By means of (pseudo) random numbers and an equilibration$^{12}$
one determine the disks center coordinates. On each brick could be
  written the 
coordinates of a disk's center (elsewhere the brick is ``empty'') 
if the coordinates are within the respective cell. After adding a new
and non-empty brick one checks ``in the vicinity'' among the already
added bricks if there are disk centers at a distance less than $r$.
Similar procedure was applied for percolation on aperiodic lattices 
(quasi-lattices).$^{13}$

The above procedure can not work for $\lambda=0$ and would be ineffective for
very small $\lambda$. An algorithm with more than one disk in a cell was
proposed in$^{10}$  
for the case $\lambda=0$. Here one faces
a ``data structure'' problem: up to now  a representation of data by
arrays with fixed size was supposed, but this time one could not determine 
{\em a priori} how many disks there would be in one cell...  The quest
for a  general approach to such kind of problems leads to the notion of
linked lists and the graph-theoretical representations which will be discussed
later.

\subsection{The cluster flip dynamics and the parallel algorithms}

An important stimulus for developing efficient parallel algorithms for
cluster counting became the cluster-flip dynamics for Monte Carlo
simulations of spin systems. Such dynamics were proposed first in 
1987 $^{14}$ 
(so called Swendsen-Wang dynamics) and after
that subjected to numerous
modifications.$^{16,17,18}$

Some spin variables defined on the lattice sites my form clusters of
``equal'' (e.g. parallel) spins. Two parallel neighboring spins may belong
to one cluster if an additional (probabilistic and temperature dependent)
rule is fulfilled.  Now ``the important property'' for a cluster is the
spin direction in it, not its size. 

For a sequential algorithm one can easily modify the HK76 method to
perform  cluster-flip dynamics. As usual, the simplest example is 
the implementation on  Ising models. Here clusters with spins "up" and 
clusters with spins "down" might be formed.
The vortex labels may then start (say) from three and
in the root of the cluster one can put  "-1" for clusters with spins up, 
and "-2" for "down" spins. Now all spins in a cluster may flips their direction
just by changing the root from "-1" to "-2" and  vica versa.

Such kinds of algorithms were proven very effective for large systems
at criticality,
and many efforts were devoted to developing parallelized codes.
Two classes of algorithms could be distinguished: for computers with
few but powerful nodes
and for massively parallel computers. It appears that massively
parallel computers
are suitable for small systems and computers with powerful nodes for large 
systems.$^19$

One can map the percolation  configuration to the configuration
of nodes for a computer with several thousand 
(or even hundreds of thousand$^{20}$ 
) processor nodes -- say, 
one (or several) site(s) per node. The labeling here relays on parallel
updating on all nodes. If each node carries one site the labeling can be  
described as ``local diffusion'' procedure$^{21,20}$
supposing that
all nodes were initialized with different label, on each update step each 
node checks all its neighbors and chooses the lowest label as its own for 
the next update step. This is repeated until no more label changes occur,
and so, nodes with equal labels form the clusters. Some further discussion 
for application to spin models can be found in ref.$^{19}$.

The algorithms for parallel computers with few but powerful nodes attract
more attention, at least because each cluster of networked workstations
may play the role of parallel computer. In that case the parallelized 
versions of
the HK76 algorithm are of prime interest.
It is accepted now that the HK76 algorithm can be successfully 
parallelized.$^{22,23}$
The idea is to divide the (square) 
sample into strips: one strip per node. 
On each strip runs (in parallel with others) a sequential HK76, keeping
the labels for the boundary vortices. Then a master node collects the data and
recovers the whole configuration. 

\section{Graph-Theoretical approaches and spanning trees}
\label{G-TA}
In a HK76-like method we start with the assumption that some
objects exist in space and a rule is known to determine which pairs of them
are directly connected (bonded). This changes a bit  when
a graph-theoretical 
framework is used.

It is assumed that a list of (abstract) elements could be
composed. Each element 
has a labeling number and each pair 
of elements might be ``bonded''.  The procedure for picking up the elements
and finding the bonds between them  is a question of technique usually 
skipped in such consideration. Here follows only a short description
of the overall procedure  (as one can  deduce it form the previous sections):

\begin{enumerate}
\item Add a new element (object, vortex..) to the list. Each element could
carry some additional information (e.g. spatial coordinates), including some 
labels eventually subject to further change.

\item Check with all (added to the moment) elements in the list, 
which of them are connected to the new one.
\item Attach to the new element a sublist which contains the numbers
  of its neighbors --
or (more generally) -- supply this element with {\em pointers} 
to its neighbors.
Update accordingly the sublists attached to each of the neighbors.
(In that way one counts a bond twice, otherwise the graph will become 
``artificially'' directed.)
\end{enumerate}

In that way we obtain a computer presentation of a {\em graph} with its
vortices (the list elements) and edges (the bonds). (For case of 
simplicity   we consider  undirected single edges --- with one
exception --- the pebble game algorithm, which is discussed in Sec. \ref{pbg}.)

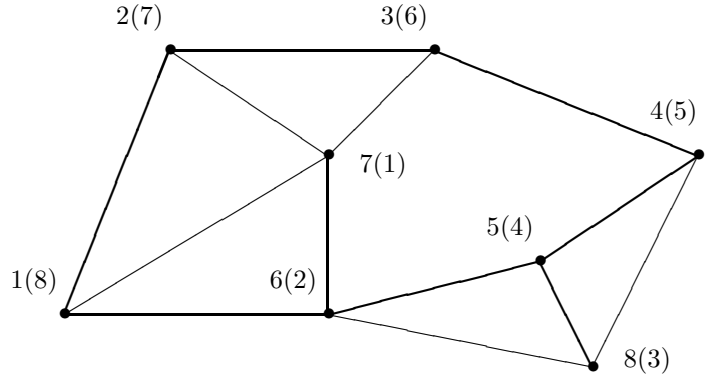
\begin{figure}
\label{DFS}
\setlength{\unitlength}{2pt}
\newcommand{\rput}[4]{\begin{picture}(0,0)(#1,#2) \put(-9,8){#3(#4)} 
                      \end{picture}}
\begin{picture}(100,80)(-21,-20)  
\put(-1,-1){$\bullet$} \put(-10,5){1(8)}
   \thicklines     \put(0,0){\line(2,5){20}} \thinlines
                   \put(0,0){\line(5,3){50}}
             \put(0,0){\line(1,0){50}}

\put(19,49){$\bullet$}  \put(10,55){2(7)} 
    \thicklines      \put(20,50){\line(1,0){50}}
    \thinlines       \put(20,50){\line(3,-2){30}} 
  
\put(49,29){$\bullet$} \put(56,27){7(1)}
                      \put(50,30){\line(1,1){20}}
  \thicklines          \put(50,30){\line(0,-1){30}}
\put(49,-1){$\bullet$}  \put(39,5){6(2)}
   \thicklines      \put(50,0){\line(4,1){40}} 
   \thinlines       \put(50,0){\line(5,-1){50}}

\put(69,49){$\bullet$} \put(60,55){3(6)}
   \thicklines       \put(70,50){\line(5,-2){50}}

\put(119,29){$\bullet$} \put(111,37){4(5)} 
   \thicklines       \put(120,30){\line(-3,-2){30}}
   \thinlines        \put(120,30){\line(-1,-2){20}}
\put(89,9){$\bullet$}  \put(80,15){5(4)}
    \thicklines       \put(90,10){\line(1,-2){10}}  
                     
\put(99,-11){$\bullet$} \put(106, -10){8(3)}
\end{picture}

\caption{An example of numbering of sites by {\em depth-first search}.}
\end{figure}

A chain of edges (chain of bonds) is called a path. A set of
vortices connected with path(s) is a connected component (a cluster) in a 
particular graph. There are  subsets of paths in a cluster, such that
the vortices remain connected, but there can not be cyclic routes
on such a subset. The name for these subsets of paths is {\em spanning tree}.
 Different spanning trees could be determined for a cluster, but
only one might exist at a time.

The well-known forest-fire model$^{2}$
gives an example
how a spanning tree can be constructed. 
(Indeed spanning tree and {\em forest tree} came from different
terminologies so the similarity is a little misleading:
Here a {\em forest tree} exists on
each vortex, and the bonds between vortices may form a {\em spanning tree}.)
As a first step a certain tree in ``the forest''  catches fire.
It ignites all its direct neighbors and burns out. At this step connections
are drawn between the first tree and all its direct neighbors.
At the next step, each burning tree ignites its ``green'' neighbors
and bonds are drawn to the ``just ignited'' trees from the tree which
ignited them.
These trees then ignite {\em their} neighbors, and so on.

The set of bonds which remains after all {\em forest trees} are burnt out is
a branched structure which connects all vortices and no cyclic routes
exist --  a {\em spanning tree} for that cluster. Different
spanning 
trees might be constructed depending on the order in which the burning
trees 
are visited on each step. 

This method of burning was applied for determining some fractal
properties of the critical (the incipient) percolation cluster$^{24,25}$
on square and cubic lattices.

Without the use of graph representations and linked lists algorithms, the above
procedure could be  quite time consuming:

One has to scan a two- (or three-) dimensional array for the square 
(or cubic) lattice which contains the studied cluster. The number of scans
is equal to the number of steps (the number of burning tree generations) in
the above method of ``burning''. The number of step ($N_s$) in its turn is 
proportional to the so called shortest path in a cluster:$^{25}$

$$
N_s \propto \ell \propto L^{d_{min}}
$$

\noindent
where $L$ is the system size, $\ell$ is the shortest path, and the 
fractal dimension of the shortest path is $d_{min} \approx 1.13$ in 
two dimensions, and $\approx 1.34$ in three dimensions. 

Another way, based again on fixed-size-array representation, is to
perform ``walks'' through the 
array-elements, with all drawbacks of the ``ant in the labyrinth'' algorithms.

Describing the same type of algorithms in a
graph theoretical framework will provide the basics for our further
considerations:

Let us  return to the enumerated  list in the beginning of this section.
It describes the procedure for the construction of
a linked-list data structure. It does not depend explicitly on the 
underlying periodic or random lattice, neither on the spatial dimensionality.
The size of the data structure is proportional to the number of elements 
irrespective of the spatial extent of its geometric counterpart.

While the case specific issues are concentrated in item 2) in the mentioned 
list, the general
problem for data links management should be solved in the last item.
If there is a maximum number of connections ($C_{max}$) for a system 
(say, with $N$ elements), a crude solution is  to create an array
with size $N \times C_{max}$ for handling the links. 
Indeed the better way is  to use based on {\em pointers} algorithmic
constructions as {\em stacks} and {\em queues}, whose descriptions
could be found in any textbook on C or PASCAL (and now FORTRAN90). 
In the following 
do not refer to any specific scheme for maintaining data links.

The point is that one can rely on well established procedures for
following the links (the connections) between elements. Two of the
most important methods for  searches on connected subgraphs$^{26}$ 
are
the {\em depth first search}\/(DFS) and the {\em breadth first search}\/(BFS).

Those methods will be the subject of two examples. First, a DFS function 
will be applied to cluster size statistics on a percolation model -- the same
task for  which the HK76 method is usually used.

The algorithm could be divided to two steps:

\begin{enumerate}
\item Create a linked list with the vortices (and the edges) 
of the studied percolation structure. Each element should ``carry'' a
Boolean 
variable "checked" set to "FALSE".

\item Pick an element of the list which is "not checked". Set to one
  the variable which will contain the cluster size ({\tt clust(i)=1})
Perform a DFS
procedure which counts the number of direct and indirect neighbors of
the picked element, mark them as "checked". Store the result. Pick the
next "not checked" element...

\end{enumerate}

In this way the data structure examination remains independent  from
its creation. The implementation of DFS  for cluster size statistics
 can be described (in a quasi-formal\footnote{The notation used here do
   not follow  strictly certain programming language, but hopefully, 
is clear enough.}                 way) as follows:

\begin{tabbing}
 {\footnotesize line 1}\=jj\=ii\= aaaaaaa\= \kill
 {\footnotesize line 1}\>\>{\bf function} DFS(vertex.V,i);\\
{\footnotesize line 2} \>\>  \>checked.V:=TRUE;clust(i):=clust(i)+1;\\
{\footnotesize line 3} \>\>  \>{\bf for} vertex.W {\bf in} (neighbors
of V): {\bf if} not checked.W {\bf do} DFS(vertex.W,i);\\
{\footnotesize line 4} \>\>  \>{\em ... the postvisit instructions
  here ..}\\
{\footnotesize line 5} \>\>{\bf return};  (end of the recursive function DFS...)
\end{tabbing}

\noindent
where "checked.V" a boolean variable included as element of a record 
 (a structure) attached to the list element "V". 
 
The whole program (including the creation of the linked list) would not 
exceed one page, and, more important, the logic of the program relies
only on the five lines here. 
( Instead of {\em recursive} function (a function which can call itself)
one can use$^{27}$
the {\em stack} type of data structure to perform
the same graph search.) The algorithmic context for the program instructions 
is  very important in general:
 the  instructions
before the line 3  
are  ``previsit'' instructions, and these after line 3 -- 
as ``postvisit''.$^{26}$

Indeed the label "checked" can hold more than two different values, so
the separated clusters may have different labels after completing the
cluster statistics. Additional labels could be defined as well; 
 clusters  could be moved to other linked lists... One can extract
 easily the largest cluster and then to apply e.g. a linked-list
 implementation of the ``burning'' algorithm mentioned above.
 
 Here follows the second example: 
the ``burning'' algorithm will be implemented by means of
the {\em breadth first search}\/(BFS) procedure. The task is to find 
the shortest path between two sites (say, "1" and "2") in a cluster. 

\begin{enumerate}
\item Create a linked list with the vortices (and the edges) 
of the studied cluster. Each element should carry {\em two} boolean variables:
"checked" and "InTheQueue", both set to "FALSE", 
and an integer variable("Level") set to $0$.

\item Attach labels to two ``opposite'' vortices
(say, "1" for the vertex with the largest $y$-coordinate and "2" for
the vertex with the smallest $y$).

\item Start from vertex "1" a {\em breadth first search}\/(BFS) for 
the element with label "2".  When the element with label ``2'' is
reached,
the  BFS could return the length of the shortest path between ``1''
and ``2'',  which is proportional 
to the level of search reached at that moment (see below and
ref.${^25}$ ). 

\end{enumerate}

\begin{figure}[htb]
\label{queue}
\setlength{\unitlength}{2pt}
\begin{picture}(100,85)

\put(10,10){\begin{picture}(100,60)(-10,-20)
\thicklines
\put(-1,-1){$\bullet$} \put(-10,5){1(0)}
                   \put(0,0){\line(2,5){20}} 
                   \put(0,0){\line(5,3){50}}
                   \put(0,0){\line(1,0){50}}

\put(19,49){$\bullet$} \put(10,55){2(1)}
                     \put(20,50){\line(1,0){50}}
    \thinlines       \put(20,50){\line(3,-2){30}} 
  
\put(49,29){$\bullet$} \put(56,27){3(1)}
                      \put(50,30){\line(1,1){20}}
                      \put(50,30){\line(0,-1){30}}

\put(49,-1){$\bullet$}    \put(39,5){4(1)}
   \thicklines       \put(50,0){\line(4,1){40}} 
                     \put(50,0){\line(5,-1){50}}

\put(69,49){$\bullet$}    \put(60,55){5(2)}
   \thicklines        \put(70,50){\line(5,-2){50}}

\put(119,29){$\bullet$}   \put(111,37){8(3)}
   \thinlines        \put(120,30){\line(-3,-2){30}}
                      \put(120,30){\line(-1,-2){20}}

\put(89,9){$\bullet$}    \put(80,15){6(2)}
                     \put(90,10){\line(1,-2){10}}  \thicklines
                     
\put(99,-11){$\bullet$}   \put(105,-9){7(2)}
\end{picture}}

\put(10,3){\dashbox{1}(8,8){1}}
\put(18,3){\dashbox{1}(8,8){}}
\put(36,0){\vector(-1,0){10}}
\put(36,3){\framebox(8,8){2}}
\put(44,3){\framebox(8,8){3}}
\put(52,3){\framebox(8,8){4}}
 \put(68,14){\vector(-1,0){10}} 
\put(68,3){\dashbox{2}(8,8){5}}
\put(76,3){\dashbox{2}(8,8){6}}
\put(84,3){\dashbox{2}(8,8){7}}
\put(92,3){\dashbox{2}(8,8){8}}
\end{picture}
\caption{An example of numbering of sites by {\em breadth first search}. 
Below is given the {\em queue} needed for the second step.}
\end{figure}
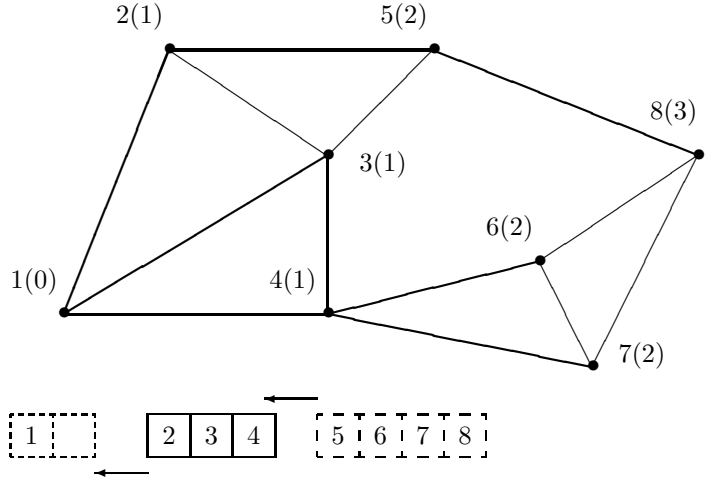

In order to describe the BFS function one needs first a description for the
{\em queue} type data structure. This is a linked list, which works
on the principle {\em first-in, first-out}, like the flow through a pipe 
(see Fig.\ref{queue}). It
suffices for the proper use of a {\em queue} if one is able to add the
``last'' element({\em E})
of the queue (the respective function will be denoted in the examples
as: {\bf Inject}({\em queue},{\em E})), 
to read the first element ({\bf Read({\em queue}})), 
to remove it ({\bf Pop}({\em queue}) ).

The BFS function could be implemented for finding the shortest path in 
a cluster as follows: 

\begin{quotation}
\begin{tabbing}
aaa \= aa \= aaa\= aa\= \kill
 \>{\bf function} BFS(vertex.V);\\
 \>  \>create a {\em queue} with the only element V; InTheQueue.V:=TRUE;\\
 \>  \>{\bf while} {\em queue} not empty\\
 \>  \> \>  $\cal V$\/:={\bf Read}({\em queue}); {\bf Pop}({\em queue}); \\
 \>  \> \>  checked.$\cal V$\/:=TRUE;\\
 \>  \> \> {\bf for} vertex.{\em W} {\bf in} (neighbors of $\cal V$\/)\\ 
 \>  \> \> {\bf if} ((not InTheQueue.{\em W}) {\bf AND} (not checked.{\em W}))
                                 {\bf do } \\
 \>  \> \> \>Level.{\em W}:=Level.{\em V} + 1;\\ 
 \> \>  \> \>InTheQueue.{\em W}:=TRUE;\\
 \> \>  \> \>{\bf Inject} ({\em queue},{\em W});\\
 \> \>  \> {\bf end do};\\
 \>  \> \> {\bf end for};\\
 \>  \> {\bf end while};\\
 \>{\bf return};  ({\em end of the function BFS...} )
\end{tabbing}
\end{quotation}

A program flow for this function is illustrated for the graph on
Fig.3. The numbers in
parenthesis correspond to the  variable ``Level'' attached to that
site. The other number corresponds to the order in which the site
``enters'' the queue. 

The solid line boxes below present the queue content before the
second execution of the instructions {\bf Read}({\em queue}) and {\bf
  Pop}({\em queue}). The arrows show the changes in the queue content
before the third execution of the mentioned instructions.

In order to compare with the ``natural'' description of the burning method,
one should point out that the instruction "checked:=TRUE" means: the 
forest-tree
"{\em V}" was ignited  in the previous step(``Level'') and will ignite its
neighbors (which are not yet burning). And, before {\bf Inject}-ing a
burning forest-tree 
into the queue, one writes down at which step this tree was ignited.
A forest-tree for which both labels "InTheQueue" and "checked" are set to
"TRUE" is an already-burnt-out one.

After completing the procedure, the integer "Level" written on each element
denotes the shortest path from the respective vortex to the vortex, ``where
the fire was initiated'' -- the starting point of the {\em
  breadth-first search}.
If the search started from vortex labeled "1"(see above)  the 
value of
"Level"
attached to the vortex labeled "2" would give the shortest path between these
two vortices (supposing all connections have equal length).

Repeating the same procedure from "2" can give other important 
characteristics:
the vortices which form so called ``elastic backbone''$^{24}$
 --- the union of the shortest paths between "1" and "2". Now we
suppose that each vortex(the respective list element) carries a second integer
label, say, "BackLevel" (keeping labels "Level" with its previous values).
After completing the second(from "2" to "1") search, the vortices for which:
$$
"Level" + "BackLevel" = TheShortestPathLength
$$
 form the ``elastic backbone''.

 The next step in the determination of the cluster structure could be 
the extraction
 of the geometric backbone . (``Almost'' synonyms for {\em geometric backbone}
 are:  {\em current carrying part} (but see  ref.$^{28}$ ),
 or,  {\rm geometric backbone} .)

 Just to present another example we will stick
 to the definition: a vortex belongs to the geometric backbone 
(in this example)
 if it lies on a self avoiding walk between "1" and "2". Indeed the geometric
 backbone includes the elastic one. We will accept that the elastic
 backbone
sites are ``black'',  the rest of backbone sites are ``gray''. All
sites
are ``white'' in the beginning. After the end of the search,  the
sites which remain ``white'' belong to the ``dangling ends'' of the percolation
structure.

The use of a label ``tested/(not tested)'' is not enough here, one could
use instead a ``search
number'' which is increased after each 
depth-first 
search. A site is ``not
checked'',
in the context below, if the ``search number'' ascribed to it is less
than the current search number.  If a site is tested during the search
it accepts the current search number. (A similar approach has been used
in the implementation of the ``pebble game algorithm'' -- see below
and
ref. $^{29,30}$ .)
 The following list presents the main items in an algorithm for
 backbone extraction based on {\em depth-first search}.

\begin{itemize}
\item Find the elastic backbone
\item Pick a "black" or "gray" vortex which has at least
one "white" and "not checked" neighbor. Start a {\em depth-first search} 
through the "white" and "not checked" vortices.
\item On each step of the DFS procedure (after the first) check for "black" 
or "gray'' neighbors. If one  is 
found, this branch of the search terminates and all sites which will
be traversed   during the
``postvist'' instructions, on the way back the origin (the vortex
which 
initiated the 
search)  have to be set gray. 
\item If the {\em depth first search} failed to find a gray or black
  site;
{\em repeat it} using as the current search number a very large number
(say, greater than the number of particles in the system). This is a
way
one to mark the sites, which do not belong to the backbone, and, hence
there is no need to be searched again.

\end{itemize}

The backbone extraction is completed 
when no more site eligible for search have remained.

\subsection{Which is better?}
 The sequence of ``burning'' examples in the previous section was
 aimed 
to show how 
 conceptually transparent 
 such algorithms could be if described in a Graph-theoretical
framework.
 One must first  adopt the spanning tree concept 
which might be difficult for the beginner.
Other 
drawbacks exists as well, but we start form the advantages of the 
graph theoretical approaches based on spanning trees:

First, one can divide the cluster counting and/or cluster description
problem to loosely connected parts, each part to be solved separately
and (almost) independently from the others. Each part could split in turn
in several, well separated subunits.  On the higher level 
one can distinguish
i) the task for creating a proper data structure -- the linked list of objects,
which clustering is studied; and ii) the cluster counting itself.

When the data structure linked list is constructed, subtasks arise as: 
how to describe the direct connections between elements, the order of
elements, what additional information  each element should carry: 
space coordinates, number and type of labels...
Sometimes the data structure is created as a result of self-dependent
computational problems e.g molecular dynamics simulation or continuous 
Monte Carlo.
Indeed it is important that the input data structure might be subject to 
extensions, say, adding new labels to each element, or reordering them.

After having the  linked list data structure 
(i.e. having a graph representation), 
one can explore the generic methods of graph searches and spanning trees.
A simple set of primitive searches could be used at that stage: combinations
of {\em depth-first search} and {\em breath-first search} functions can solve 
in a transparent and effective way many tasks related to the cluster
separation and the cluster structure. 

The computational complexity of these searches is 
usually
known, so the
computational costs could be predictable, and well planned. Usually the
efforts rise linearly with the number of elements, the problem is
that the proportionality coefficient is usually large. Here start the 
drawbacks of the linked-lists. 

As a rule one must keep in the central memory the whole structure
which is in contrast with the HK76 algorithm where only (d-1) dimensional
cuts are kept together with the Labels-of-Labels array.
Sometimes one must keep the whole structure even with HK76 type of algorithm, 
but the memory requirements for linked-list approaches might remain higher by 
factor up to ten, compared with HK76 and other methods based on 
fixed-size-arrays representation.

The depth of the recursion stack is not infinite, and one should
keep in mind its possible overflow for large structures.

Perhaps the most serious problem: one can hardly imagine effective 
implementations of parallel codes for a linked list data structure 
and for recursive searches on it.  Instead, one could use the {\em trivial 
parallelization}
for statistical problems -- to run independent ensemble samplings on the
different processing nodes.

The contradiction could be expressed as: a spanning three
approach may perform faster for large systems, but needs more memory
and can not be parallelized, so, very large systems are inaccessible to it.

Some time  combined methods which use both a HK76 scheme and a spanning 
tree searches might be the best choice. 
 A possible combination could be sketched as follows:.

A very large ($L \times L$) percolation system is studied by a
parallel 
computing system: say, a cluster of $(N+1)$ workstations -- 
$N$ nodes and a host. The sample is divided on $L/N \times L$ strips. 
It is {\em sweeped} three times. The first sweep is a parallel
version
of a HK76 algorithm (e.g. as in ref.$^{22}$ ). 
A second sweep of
the parallel  code with the same sequence of random numbers would
allow
to extract the coordinates of  some of the percolation clusters.
On the third step clusters these clusters are distributed to the
nodes;
so each node could perform spanning tree searches on a different cluster.
So, in such a combination
the Hoshen-Kopelman algorithm would work as a kind of preprocessor for
the spanning tree searches.

\subsection{A final example: the pebble game in two dimensions}
\label{pbg}

We have considered up to now only  simple connectivity clusters and
undirected graphs. Here we will present  a case where a non-local
condition for connection between  sites forming a cluster is imposed.
It will lead us to search on directed subgraphs.


One considers subclusters to the usual connectivity percolation
clusters imagining that each bond represents a rigid bar and each site
a (flexible) joint. The task is  to recognize the subset(s) of
bonds(bars) and sites(joints) which are mutually rigid. It means: if one
applies a stress between any two sites which belong to a rigid subset
 there is no possible movement of the sites without changing the
 length of some of the bonds in this subset.

The task could be solved in a very elegant way by using an analogy
taken from the classical mechanics of point masses. As known the
total number of degrees of freedom($F$) for a system of $N$ point masses in $d$
dimensions is simply $F= d.N - C$, where $C$ is the total number of
{\em independent} restrictive conditions. In our case each point mass
is a site (a joint) and each constraint is imposed by adding a bond(rigid bar),
between given two sites. So, a rigid (sub)cluster corresponds to a
{\em rigid body} in the sense of classical mechanics:  such cluster
should
have  three degrees of freedom only (in 2d): they
correspond to two
translations and one rotation of the whole set of point masses.

The task is reduced to answering the question: which constraints are
independent? The way of getting this answer could be described again
by a build-up analogy: this time instead bricks each time new bar is
added and a decision is taken if this bar impose a new constraint or
it is redundant for the percolative rigidity.

Now a ``pebble game'' $^{29,30}$
can be introduced:
we start with the sites of an unconstrained lattice (point mass
system). To each site(point mass) are attached two free pebbles 
which correspond to the two degrees of freedom of the  point mass.
The ``pebble game'' starts with adding the first bar (present bond)
between two (neighboring) sites. Then one of the sites is chosen at
random
and one  of ``its'' 2 free pebbles is moved to
``cover'' the bar: it becomes ``anchored'' to the closer end of the
bar. In this way a direction is defined: say, pointing
outward from form the ``anchor'' end. 
The site on the anchor and still
has two pebbles but one of them is ``anchored'' and the other is
``free''. So the system of two joints and a rigid bar connecting them
has now three  degrees of freedom. and represent a directed subgraph with
two sites and a directed bond between them. Adding each new bond one has to
check if there are two free pebbles  on each of the sites ``under
connection''. If one of the sites has less than two
free pebbles then the  respective site belongs to a directed subgraph
and one or both of its pebbles are anchored to (a) previously added bond(s).

Now one has to mention that in directed subgraph one can change the
direction (preserving the attachment of  two ``initial'' degrees of
freedom to each site)  of a bond if there is at least one free pebble 
on the site to which the bond is pointing. Then the free pebble can be
anchored to the opposite end freeing the previous anchor. This changes
the direction of the bond and moves a free pebble from one site to its
neighboring site. Using such elementary steps one can shuffle the free 
pebbles along Cretan paths in the subgraph.

The aim of the ``pebble game'' is always to ensure four free pebbles
for the two sites to which a new bar is added. Three pebbles could be
always ensured. If the search for the forth failed this mean that the
new constraint is not independent. The prove for this could be find
elsewhere $^{31}$ 
we will start the description of  pebble game 
implementation based on {\rm depth-first search} in the directed subgraphs.

It is supposed that  all the sites in the lattice are (consecutively)
numbered, and each line in the input file consists of 
two numbers -- the numbers of sites which share a bond. A proper
linked-list structure for the lattice sites should be  designed as well. 
Here, the neighbors sublists attached to each site should be 
enlarged with a label for each ``out-coming'' bond: in case
 a pebble is attached to it or not. The number of bonds with pebbles
 for this site can be found after inspecting these labels. Now the free
 pebbles 
on a site are just  \ $2 - {\tt TheNumberofBondsWithPebbles}$.

\noindent
The following description
could show what a proper structure should be.

\begin{enumerate}

\item Read a bond. Add each of the two sites into the linked list, if
  not already added. If one (or both) of them are already in the list,
  rearrange the neighbors list for  this(these) site(s).

\item Try (by a DFS function -- see below) to free two pebbles on
  both sides of  the bond (all together 4 pebbles)

\begin{enumerate}

\item Check how many free pebbles belong to the two sites.
If this number is 4 attach one of them to the new bond.
 
if not:

{\bf *} Pick one of the bonded sites (say, $n_1$):

\begin{itemize}

\item {\bf if} not found 2 free pebbles on it 
             {\bf do} DFS($n_1$);

\end{itemize}

{\bf *} Pick the second end site ($n_2$) and repeat the same 
(... DFS($n_2$);.. ).

\item Repeat from item 2; until 4 pebbles are found, or until 
the search originated from both sites failed

\item If the search failed to free 4 pebbles, the new bond is
redundant (in view of rigidity).

\end{enumerate}

\item Repeat starting from 1, reading new bonds ...
\end{enumerate}

The application of the {\em depth-first search} algorithm in the
``pebble game'' requires one
argument -- the site for the next visit.  The DFS({\tt s}) function itself 
returns a boolean value: free pebble found or not (DFS({\tt s})= TRUE or
FALSE).
The function traverses this time a directed graph. The ``out''
direction
starts from the bond end where a pebble is attached.

The  {\em previsit instructions} here consist in:  

\begin{quotation}
Check {\bf if} there is at least one free
pebble on the visited site,\\
 if yes {\bf then}:

DFS:= TRUE; {\bf exit} from this instance of the recursive function...
\end{quotation}
 
If  a free pebble is not wound during the previsit, 
then a new site for the next search have to be chosen:

{\bf Peek} a neighbor of {\tt s}, say, {\tt os}. {\bf If} site {\tt os} has not been
visited this search (tag({\tt os}) $<$ searchtag) {\bf and} the bond
 is directed outward ({\tt s}---$>${\tt os}) {\bf then} call
 recursively the same function (DFS) with argument {\tt os}. This call
 could be realized as an assignment:

  {\tt FoundFreePebble}:=DFS({\tt os});

If {\tt FoundFreePebble} is TRUE; during the postvisit instructions an
exchange is made: The pebble which belong to {\tt s} and was attached
(``anchored'' $^{29,30}$
to the bond {\bf\tt s}---{\tt os} now
becomes
free and the previously free pebble of {\tt os} is attached to the
traversed bond from the {\it os} side,
this way changing its direction:  {\tt s}$<$---{\tt os}. 
This is repeated recursively until
the
free pebble reaches the site where the search was originated.



This completes the description of  a {\em depth-first search}
implementation for counting of redundant constraints in a rigidity percolation
model.
The whole example program (about 60 lines of text) could be obtained 
upon request from the author.

The program could be modified  for counting of the rigid
clusters if one keeps in mind that a site may belong to more than one
rigid cluster $^{29,30}$
and that two sites belong to one
 cluster if they share a bond or if a new (long range bond) added
between them appears redundant. The bad news is that the
performance scaling of this program is $\approx N^\varpi $ with $N$ the
number of sites and $\varpi \approx 1.8 - 2$. One can improve the algorithm
performance scaling to $\varpi \approx 1.1 - 1.2$ by ``artificially''
changing the connections between sites: either by
``triangularization''$^{32}$
or by introducing of supersites (``rigid bodies'' -- with three degrees
of freedom each
--- see ref. $^{33,34}$ ).
 So, the plain use of spanning tree
searches
is not a guarantee to develop a fast algorithm.

{\em In conclusion} I would stress again that in my opinion it is very
important
one to be awared for the general properties of these two approaches
to the cluster counting problem.  I hope that this work might
show what should be expected from a program for cluster counting and
what kind of improvements or modifications  could be looked for.
So, for every particular case one can
chose the more suitable approach, or to look for a combination between
them. Generally said the Hoshen-Kopelman algorithm is more suitable
for  very large systems (say, above $10^7$ or $10^8$ ``particles'')
and is easy for parallelisation. The spanning tree approaches can
solve easier more sofisticated tasks (e.g. for percolation backbone
extraction, or for identifying rigidity percolation clusters) on
systems with moderate size.

\nonumsection{Acknowledgments}

I thank  to H.J Herrmann and S. MacNamara for the critical reading of
the manuscript. This work was supported partially by a grant from the German
Academic Exchange Foundation (DAAD).

\nonumsection{References}

\end{document}